\documentclass{eas}
\usepackage[utf8]{inputenc}

\usepackage[sort&compress]{natbib}
\bibpunct{(}{)}{;}{a}{}{,}
\usepackage{color}
\definecolor{darkgreen}{RGB}{0,142,128}
\newcommand{\todo}[1]{\textcolor{darkgreen}{#1}}

\usepackage{hyperref}
\usepackage{wrapfig}
\hypersetup{colorlinks,citecolor=blue,linkcolor=blue}

\usepackage{graphicx}

\title{Physics of star-planet magnetic interactions}
\author{Antoine Strugarek}
\date{October 2019}

\begin{document}

\begin{abstract}

Magnetic interactions between a planet and its environment are known to lead to aurorae and shocks in the solar system. The large number of close-in exoplanets that have been discovered so far triggered a renewed interest in understanding magnetic interactions in other star-planet systems. Multiple magnetic effects were then unveiled, such as planet inflation or heating, planet migration, planetary material escape, and even some modifications of the host star apparent activity. Our goal here is to lay out the basic physical principles underlying star-planet magnetic interactions. We first briefly review the hot exoplanets' population as we know it. We then move to a general description of star-planet magnetic interactions, and finally focus on the fundamental concept of Alfvén wings and its implication for exosystems.   
\end{abstract}

\maketitle

\section{Introduction}
\label{sec;Introduction}

\subsection{The hot exoplanets population}
\label{sec:HotExo}

\begin{wrapfigure}{r}{0.5\textwidth}
    \centering
    \includegraphics[width=\linewidth]{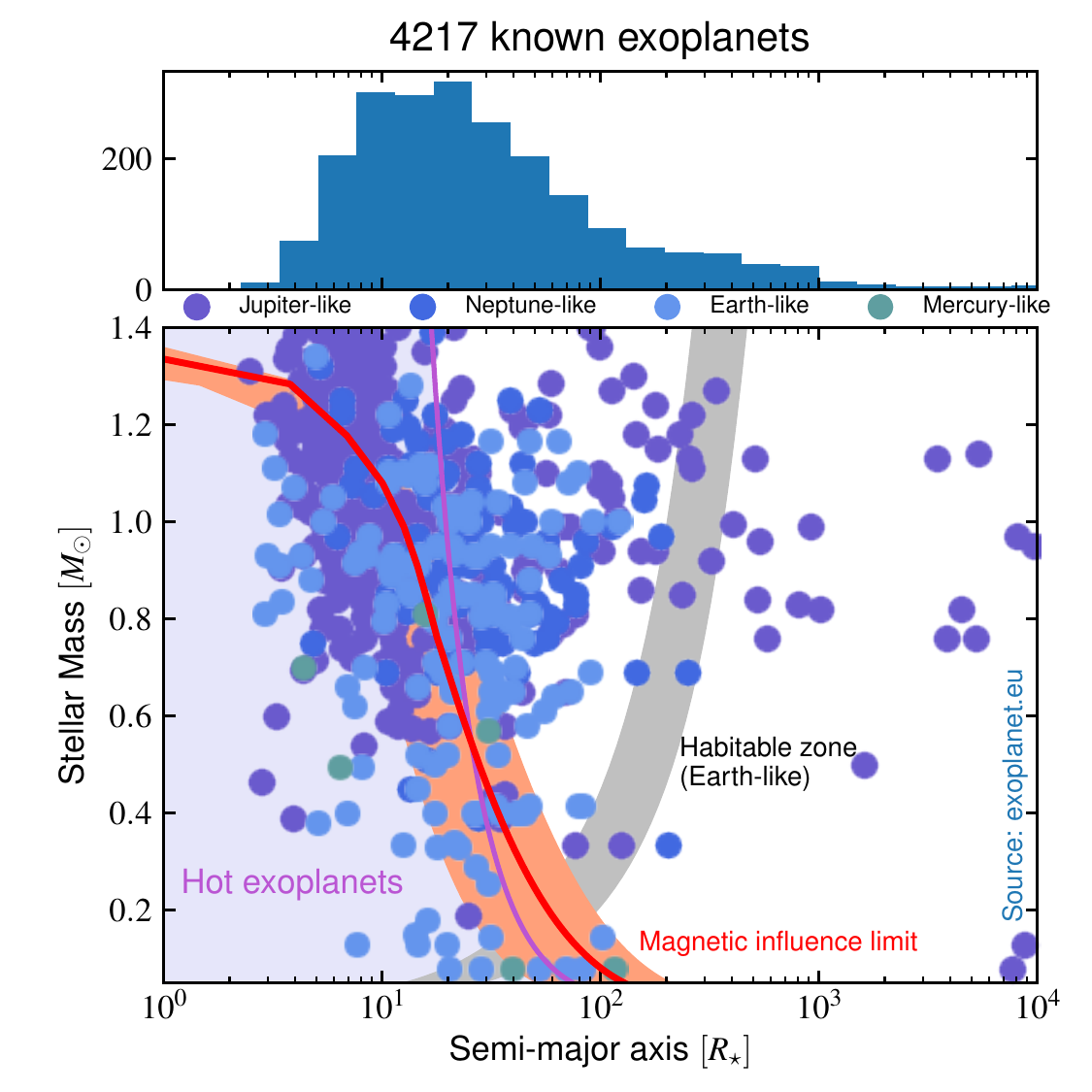}
    \caption{Exosystem distribution from \href{http://exoplanet.eu}{exoplanet.eu}. The bottom panel shows the distribution as a function host-star mass (vertical axis) and semi-major axis (horizontal axis), colored by the planet mass ($M_p\sin i$). The top panel shows the same distribution only as a function of the semi-major axis normalized to the stellar radius. The classical habitable zone \citep{Kopparapu2013} is shown as a grey band for an Earth-like planet. The $P_{\rm orb}=10$ days is labelled by the magneta line. Finally, a 'magnetic influence limit' line is shown in red (see \S\,\ref{sec:subsuper}).}
    \label{fig:ExoDist}
\end{wrapfigure}

In the long trailing wake of the discovery of 51 Peg b \citep{Mayor1995}, more than 4000 exoplanets have been detected today. Space-based observatories such as \href{http://kepler.nasa.gov}{Kepler} and \href{https://corot.cnes.fr}{Corot} have revolutionized this field by bringing a long-awaited wealth of exosystems, including unexpected massive planets on very short-period orbit (the so-called \textit{hot Jupiters}). Complementing transit and radial velocity detection techniques, ground-based coronography further allowed the direct imaging of distant exoplanets and revealed the existence of planets on very wide orbits \citep[for a review, see][]{Pueyo2018}. The heterogeneous population of detected exoplanets is shown in Fig \ref{fig:ExoDist} as a function of the hosting star mass (vertical axis) and the orbital semi-major axis (horizontal axis), in units of stellar radius. The top panel presents the aggregated histogram of the known exoplanets as a function of their semi-major axis, which is heavily biased towards planets on compact orbits with a semi-major maxis typically smaller than 20 stellar radii\footnote{Note that in the solar system, the semi-major axis of Mercury is about 83 solar radii}. Such close-in planets are expected to strongly interact with their host star through a variety of physical processes: they receive a strong irradiating stellar flux that affect their upper and lower atmospheres \citep{Gronoff2020}, they are subject to intense tides \citep{Mathis2018a}, and they orbit in a strongly magnetized medium. We will focus here on the latter and refer the reader to the cited reviews for the other types of star-planet interaction.

\subsection{Observational status of star-planet magnetic interactions}
\label{sec:HotExo}

The search for star-planet magnetic interactions has been a continuous history of positive and negative detections. The first firm detection of a signal that could be related to star-planet magnetic interactions was arguably reported by \citet{Shkolnik2003}, following a seminal theoretical argument for the existence of star-planet magnetic interaction by \citet{Cuntz2000}. They monitored at that time 5 promising star-planet systems, and detected a modulation associated with the hot Jupiter orbital period in the K band of Ca II of HD 179949. The other systems did not show any direct evidence of star-planet magnetic interactions at that time. Such signals have been observed since for different systems (see the review by \citealt{Shkolnik2017}) such as famous compact exosystems like 55 Cnc or HD 189733 \citep{Cauley2019}. The star-planet interactions tracers found in the stellar activity indicators have always been subject to a large variability with very clear signals at some epochs, and no detection at others \citep[see for instance][]{Shkolnik2008,Cauley2018}. We believe today that at least part of this hide \& seek stems for the intrinsic variable nature of star-planet magnetic interactions. Star-planet magnetic interactions are controlled by the large-scale magnetism of stars, that can be topologically complex \citep{Donati2009,See2015} and time-variable due to eruptive events (\textit{e.g.} coronal mass ejection, on a monthly time-scale) or magnetic cycles (on the yearly to decadal time-scale). As a result, star-planet magnetic interaction likely varies in nature and strength along planetary orbits as well as longer timescales.

Historically, tracers of star-planet magnetic interaction have been searched for in radio. Indeed, compact exosystems look alike planet-satellite systems in our own neighbourhood. The interaction of Io in the magnetosphere of Jupiter is known to lead to cyclotron-maser radio emissions at the poles of Jupiter \citep{Zarka2007}. At the same location, strinkingly clear UV spots are present at the pole of Jupiter \citep{Clarke2002a}, tracing the footprints of the magnetic interaction of Jupiter with Ganymede, Io, and Europa. Until recently, no clear radio detection of star-planet magnetic interactions have been successfully obtained \citep[for a recent review see][]{Zarka2018}. In 2019-2020, two new observations hinted toward these long-awaited radio emissions. The first one was a cautious suspicion of a radio emission in the $\tau$ Boo system by \citet{Turner2019a}. The second one was the detection of coherent radio emission from the red dwarf GJ 1151 by \citet{Vedantham2020}. The latter is an indirect detection in the sense that the observed stellar radio emission is peculiarly intense compared to what one would expect from such a slowly-rotating star. The authors showed that the observed emission could be compatible with star-planet magnetic interaction with an Earth-like planet on a 1 to 5 days orbit, but we have not detected this planet so far. A firm detection of this hypothetical planet through transits or radial velocity would help strengthen our confidence in this detection scenario today. 

\section{Magnetic interactions in compact exosystems: overview}
\label{sec:Overview}

\subsection{Sub- \textit{vs} Super-alfvénic interaction}
\label{sec:subsuper}

Planets in close-in orbits are generally thought to have their orbit circularized by efficient tidal interactions \citep{Souchay2013}. We will therefore consider here only planets in circular orbit, for the sake of simplicity. The characterization of star-planet interactions requires to estimate the plasma conditions along the orbit, that are set by the stellar wind. We will not detail stellar wind theory here and refer the reader to the previous Evry Schatzmann school proceedings by \citet{Brun2020}[\todo{Johnstone, this volume}] where the basics of magnetized stellar winds are introduced \citep[the interest read can start by the seminal papers of][]{Parker1958,Weber1967}. Let us summarize it quickly as follows. Cool stars like the Sun possess an external turbulent convective envelope that is the seat of an efficient dynamo process. The associated large-scale magnetic field then sculpts the environment of the star. In addition, turbulent photospheric motions are thought to excite magnetic perturbations, likely in the form of Alfvén waves, that propagate along this magnetic field and ultimately deposit energy upper the stellar atmosphere. In the case of the Sun, this leads to a very hot corona reaching more than a million Kelvin. This extremely hot corona then leads to the existence of a stellar wind, accelerating outwards and reaching super-sonic and super-alfvénic speeds. At the Earth orbit, the solar wind speed averages between 400 and 700 km/s, and reaches an alfvénic Mach number $M_a$ between 9 and 10! In the context of exosystems, one can then wonder what is the speed of the stellar wind around close-in planets.

\begin{figure}
  \centering
  \includegraphics[width=\linewidth]{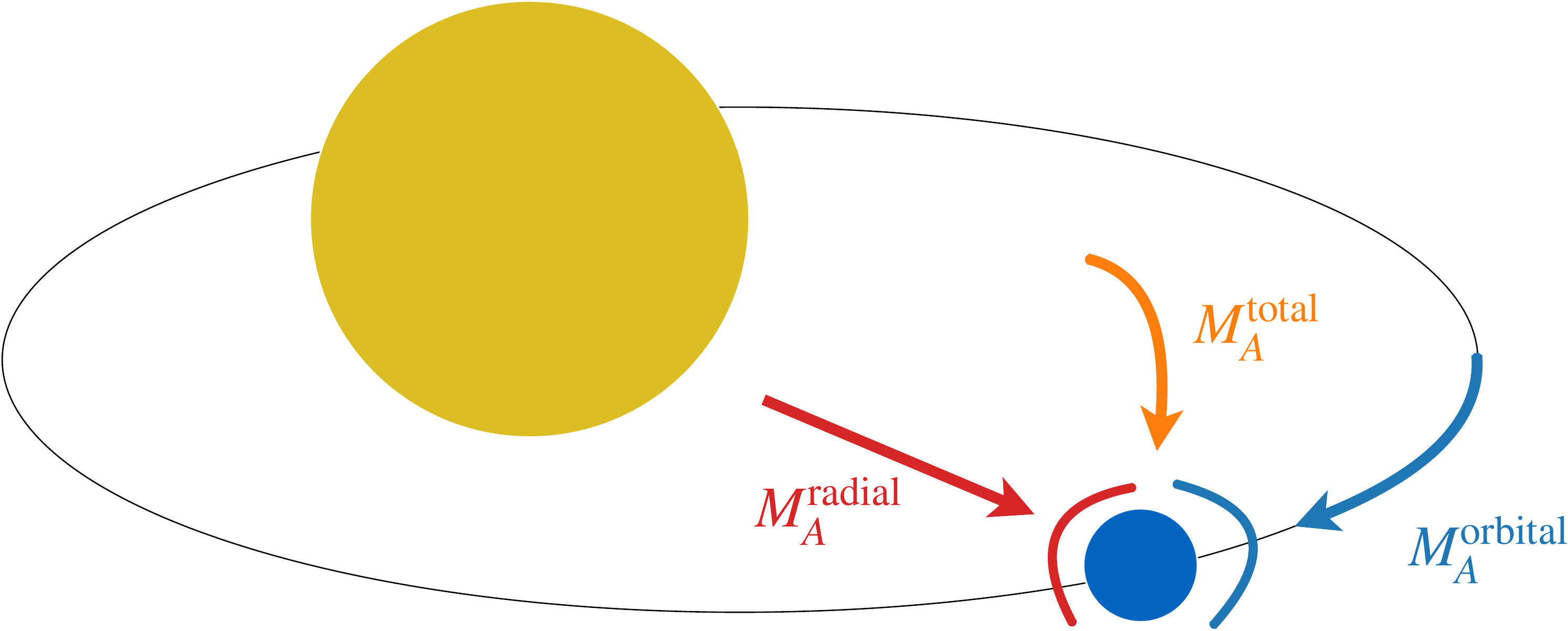}
    \includegraphics[width=\linewidth]{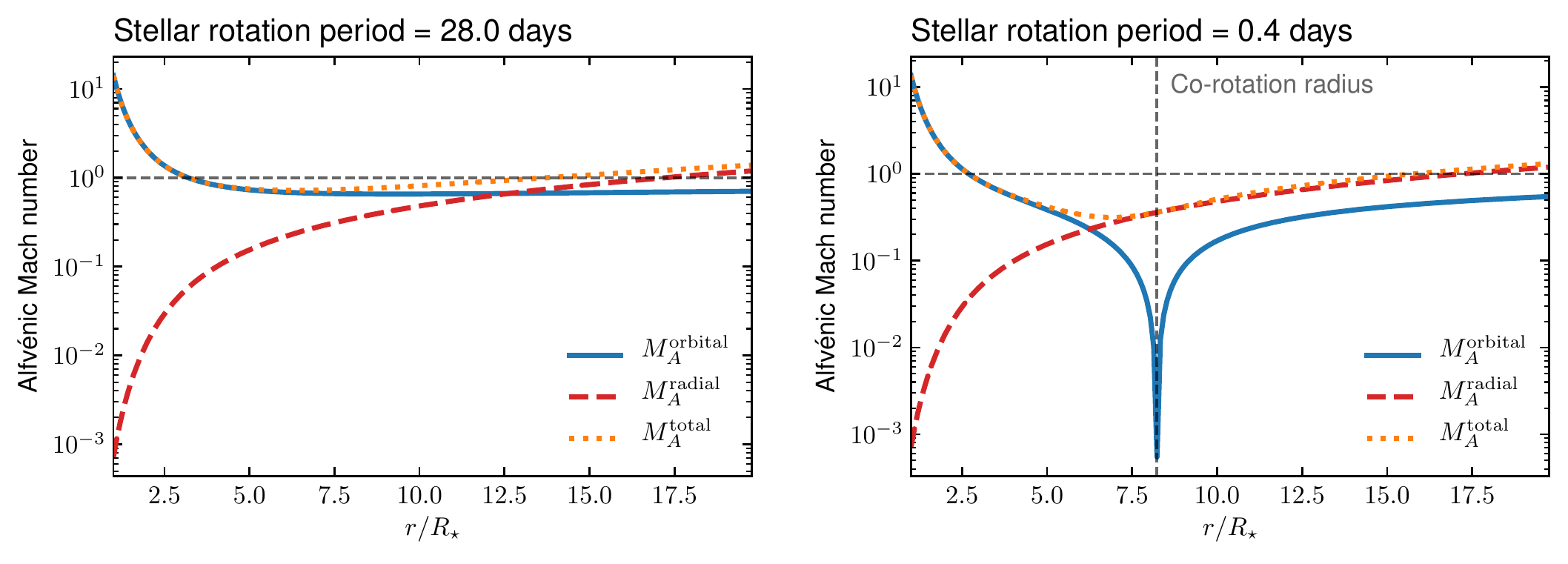}
    \caption{Schematic of alfvénic Mach numbers in compact exosystems. The top panel illustrates the star (yellow), the planet (blue) and its orbit (black), and the direction associated with the alfvénic Mach numbers (Eq. \ref{eq:MaR}-\ref{eq:MaT}). The bottom panels show the values of these numbers as a function of orbital distance. The left panel corresponds to a solar-like star, and the right panel to a young, fast-rotating Sun.}
    \label{fig:WindMa}
\end{figure}

We can get a first estimate by computing a simple Weber-Davis \citep{Weber1967} wind solution under the influence of rotation. Several alfvénic Mach numbers can be computed, each characterizing different aspects of star-planet magnetic interactions. They are summarized in Fig. \ref{fig:WindMa} and defined as follows

\begin{eqnarray}
    M_A^{\rm radial} &=& \frac{v_r^w}{v_A}\, , \label{eq:MaR}\\
    M_A^{\rm orbital} &=& \frac{\left|v_\varphi^w - v_{\rm kep} \right|}{v_A}\, , \label{eq:MaO}\\
    M_A^{\rm total} &=& \frac{\left|{\bf v}^w - {\bf v}_{\rm kep} \right|}{v_A}\, , \label{eq:MaT}
\end{eqnarray}

where ${\bf v}^w$ is the stellar wind velocity field, which we consider to be primarily in the radial star-planet direction (\textit{i.e.} the accelerating component, denoted $v_r^w$) and in the orbital direction (due to stellar rotation, denoted $v_\varphi^w$). The Kepler velocity ${\bf v}_{\rm kep}$ is assumed here to be purely in the azimuthal direction (circular orbit). Finally, $v_A=B_w/\sqrt{\mu_0 \rho_w}$ is the local Alfvén speed at the planet orbit based on the plasma properties in the stellar wind (where $\mu_0$ is the vacuum magnetic permeability).

In the bottom panels of Fig. \ref{fig:WindMa} we illustrate the different alfvénic Mach numbers for a solar-like star on the left and a fast rotating Sun (\textit{e.g.} a young Sun) on the right. The classical alfvénic Mach number, based on the radial velocity of the stellar wind (Eq. \ref{eq:MaR}), is shown by the red dashed lines. For the Sun, this simplified 1D model predicts an average alfvénic point ($M_A^{\rm radial}=1$) around 17 solar radii (\textit{i.e.} 0.08 AU). The orbital Alfvénic Mach number is quite different as it is based on the relative orbital motion of the planet in the stellar environment. At first order, the corona of a star can be approximated to co-rotate with the star below the alfvénic point, and to rotate with a profile $v_\varphi^w\propto r^{-2}$ conserving angular momentum in the super-alfvénic region outside.

The orbital alfvénic Mach number $M_A^{\rm orbital}$ is shown in blue in both panels. It is super-alfvénic only very close to the star, due to a very fast keplerian motion. We see that it remains nevertheless sub-aflvénic for most of the domain for a solar-like star. In the case of an extremely fast-rotating star (right panel), the co-rotation radius (vertical dashed line) can be in the sub-alfvénic domain of the stellar wind. At the co-rotation radius, $M_A^{\rm orbital}=0$ by definition, which means that the magnetic interaction vanishes. Note that it is nevertheless a non-stable equilibrium of the star-planet system (see for instance \citealt{Strugarek2014e}).

One can conclude from this crude analysis that many of the planets orbiting below $20\, R_\star$ are likely to be in a sub-alfvénic interaction regime, with all Alfvénic Mach numbers below 1. We illustrate this by applying our simple 1D Weber-Davis wind model to stars with varying mass and rotation rate (typically here between 2 and 28 days). We obtain an estimate the Alfvénic transition, which appears as the red area along with the exoplanet distribution in Fig. \ref{fig:ExoDist}. Note that some very close-in planets could still be in a super-alfvénic interaction state in their orbital direction, which still allows them to be magnetically connected to their host star in the radial direction. We now turn to the basic description of different magnetic connection scenarii in the sub-alfvénic regime.

\subsection{Different regimes of sub-alfvénic interactions}
\label{sec:dipunip}

A planet on a short-period orbit can be thought of as a perturber placed in a likely non-axisymmetric interplanetary medium. The perturbations it triggers can be decomposed on magneto-hydrodynamic (MHD) waves traveling away from the planet location (we will give the full description of this in \S\,\ref{sec:MHDwaves}). A net Poynting flux is therefore channeled away from the planet by waves, along what is often referred to as Alfvén wings (see \S\,\ref{sec:AlfWing}). These alfvénic perturbations can travel back and forth between the star and the planet due to reflection along the magnetic field gradients. In the context of exosystems, a simple estimate of the ratio between the alfvénic travel time and the advective time-scale of the flow across the obstacle can be found in \citet{Strugarek2018a}. In essence, what is found is that for obstacles constituted by a planet and its magnetosphere, alfvénic perturbations generally do not have time to perform a back and forth travel between the planet and the low atmosphere of the star during the advective crossing time-scale\footnote{Note that in general, the reconnection timescale is short enough to not play major role here in setting the type of star-planet magnetic interaction. It can nevertheless play a role in setting the absolute amplitude of these interactions.}. This situation corresponds to the so-called \textit{pure Alfvén wing case} \citep{Neubauer1998}. Other scenari can be achieved, depending on the propagation time of these waves and on the conductive properties of the planet. For instance, \citet{Laine2012a} showed that planets with no magnetosphere and high internal conductivities could lead to the opposite situation where Alfvénic perturbations of have time to perform multiple back-and-forth travels. The later case is generally referred to as the \textit{unipolar inductor} interaction regime. A full description of the different interactions regimes can be found in \citep{Strugarek2018a}, we refer the interested reader to this review and the references therein. While Alfv\'en wings have been observed on Earth \citep[\textit{e.g.}][]{Chane2012a} and for natural moons of solar system's giants planets \cite[\textit{e.g.}][and references therein]{Clarke2002,Bonfond2017e}, the distinction between the different scenarii has proved to be difficult observationally. In all cases, though, two Alfvén wings always form in the (${\bf B}_w,{\bf v}^w-{\bf v}^{\rm kep}$) plane. Depending on the Alfvénic Mach numbers (\S \,\ref{sec:subsuper}), both wings can connect onto the host star; only one may do so while the other extends away toward the interplanetary medium, or both may head away from the host star. As a result, the knowledge of the magnetic configuration between the host star and the orbital path of the planet is mandatory to assess star-planet magnetic interactions. 

\section{Magnetic interactions in compact exosystems: the Alfvén wings}
\label{sec:AlfWing}

We now turn to the theoretical concept behind our understanding of star-planet interactions: the concept of \textit{Alfvén wings} \citep{Neubauer1980}. 

\subsection{Pre-requisites on magneto-hydrodynamic waves}
\label{sec:MHDwaves}

Let us recall here the properties of large-scale waves in magnetized plasmas. Neglecting rotation and stratification, three MHD modes can be identified: a slow MHD mode, a fast MHD mode, and pure Alfvén mode. The associated waves will be dubbed Alfvén wave (AW), slow MHD wave (SW), fast MHD wave (FW). Note that the term \textit{magnetosonic} is generally used for purely perpendicular waves, we thus prefer to keep here the generic \textit{MHD} denomination for these modes and waves. The derivation of the dispersion relation of these modes can be found in most good MHD textbooks (e.g. \citealt{Goedbloed2019} to cite only one). We give only here the result and let the reader dig into such textbooks for their derivation.

We will write ${\bf k}$ the propagation vector of the wave, and $\omega$ its frequency. We will assume that at the studied scales the magnetic field can be considered to be homogeneous and we write ${\bf B}_0 = B_0 \hat{\bf b}$. The angle between ${\bf k}$ and ${\bf B}_0$ will be written $\theta$. Finally, the classical Alfvén and sound speeds are defined as 
\begin{eqnarray}
    {\bf v}_A &=& \frac{{\bf B}_0}{\sqrt{\mu_0 \rho_0}}\, , \\
    c_s &=& \sqrt{\frac{\gamma P_0}{\rho_0}}\, ,
\end{eqnarray}
where $\rho_0$ the plasma density, $P_0$ the plasma pressure and $\gamma$ the ratio of specific heats. The dispersion equation for a pure Alfvén wave is
\begin{equation}
    \frac{\omega}{k} = v_A \cos\theta\, ,
\end{equation}
and the dispersions for fast (+) and slow (-) MHD waves are
\begin{eqnarray}
    \left(\frac{\omega}{k}\right)^2 &=& \frac{1}{2}\left[ \left(c_s^2+v_A^2\right)\pm\sqrt{\left(c_s^2+v_A^2\right)^2 - 4 c_s^2v_A^2\cos^2\theta} \right] \nonumber \\
    &=& \frac{v_A^2}{2}\left[ \left(\tilde{\beta}+1\right) \pm \sqrt{\left(\tilde{\beta}+1\right)^2-4\cos^2\theta} \right]\, ,
\end{eqnarray}
where $\tilde{\beta}$ is related to the plasma $\beta$ through 
\begin{equation}
\tilde{\beta} = \frac{c_s^2}{v_A^2} = \frac{\gamma}{2} \frac{2\mu_0 P_0}{B_0^2} = \frac{\gamma}{2}\beta\, .
\end{equation}

\begin{figure}[htbp]
    \centering
    \includegraphics[width=\linewidth]{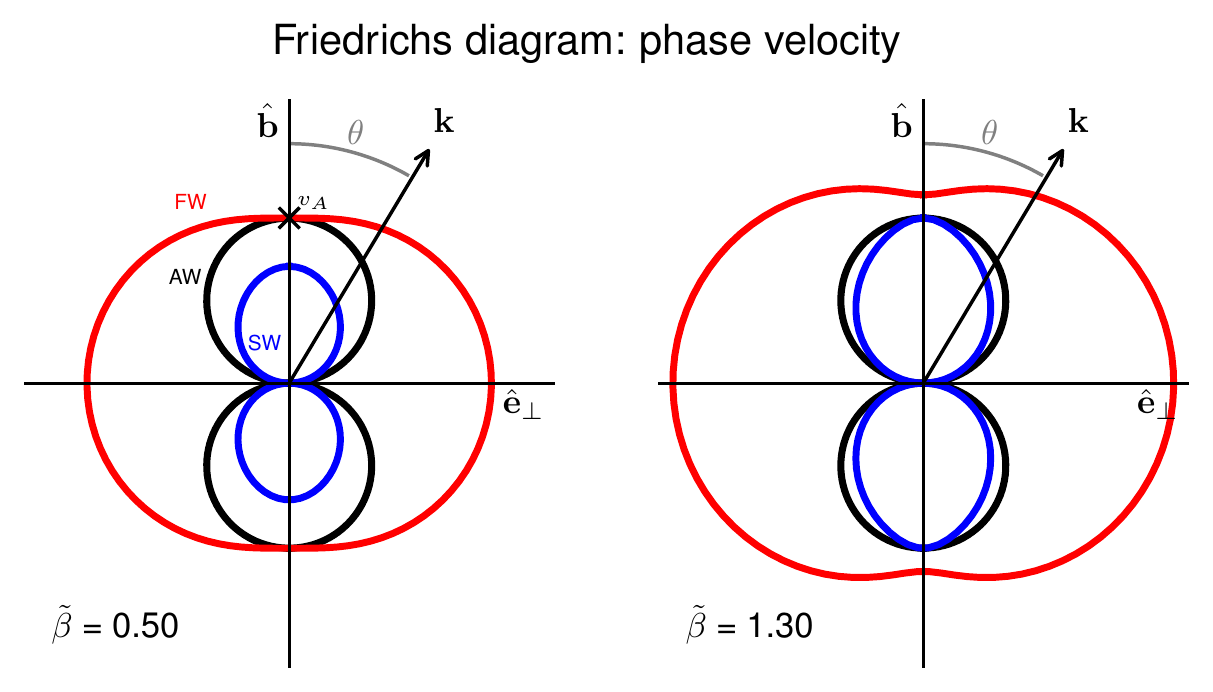}
    \caption{Friedrichs diagrams of the phase velocity of Alfvén waves (AW, black), fast MHD waves (FW, red) and slow MHD waves (SW, blue). The vertical axis is aligned with an ambient magnetic field, the horizontal axis represents any perpendicular axis. On the left panel the Alfvén speed is larger than the sound speed, on the right panel it is the opposite.}
    \label{fig:PhaseDiag}
\end{figure}

A graphical representation of Alfvén waves can help guide our intuition on their properties. In particular, Friedrichs diagrams can be built for these three fundamental waves. For a given frequency, we represent in Figure \ref{fig:PhaseDiag} the phase speed of the wave ${\bf v}_p = (\omega/k)\hat{\bf k}$ for all propagation directions $\hat{\bf k}$, where $\hat{\bf k}$ is the unit vector aligned with ${\bf k}$. In each panel, the magnetic field direction ${\bf \hat{b}}$ is represented by the vertical direction. The perpendicular direction ${\bf e}_\perp$ is represented by the abscissa. In 3D, the Friedrichs diagrams are invariant by rotation along ${\bf \hat{b}}$. As a result, we represent these diagrams in 2D, bearing in mind that the full characteristics can be obtained by rotation along ${\bf \hat{b}}$.

We represent the phase speeds for the AW (black), the SW (blue) and FW (red). The AW and SW are degenerate for ${\bf k}=0$ and fall back on what is sometimes called an \textit{entropy wave}. Two illustrative cases are shown in Figure \ref{fig:PhaseDiag}. On the left panel, $\tilde{\beta} < 1$ and the AW and FW have the same phase speed when ${\bf k} \parallel {\bf \hat{b}}$. On the right panel, $\tilde{\beta} > 1$ and the AW and SW have the same phase speed when ${\bf k} \parallel {\bf \hat{b}}$. We note that as $\tilde{\beta}$ increases, the sounds speed overcomes the Alfvén speed and the AW and SW converge essentially to the same the phase speed portrait as the dispersion relation of SWs converges to the one of AWs.

The same diagram can be build for the group velocity of the waves. The group velocity is defined as

\begin{equation}
    \label{eq:GroupVelDef}
    {\bf v}_g = \frac{\partial \omega}{\partial {\bf k}} = \hat{{\bf k}} v_p + k\frac{\partial v_p}{\partial {\bf k}}\, .
\end{equation}
The first term corresponds to the parallel component of the group velocity (which is equal to the phase speed), and the second to perpendicular component of the group velocity. In the case of AWs, we directly obtain that only the parallel component remains and obtain

\begin{equation}
    \label{eq:GSaw}
    {\bf v}_g = v_A {\bf \hat{b} }\, .
\end{equation}

\begin{figure}
    \centering
    \includegraphics[width=\linewidth]{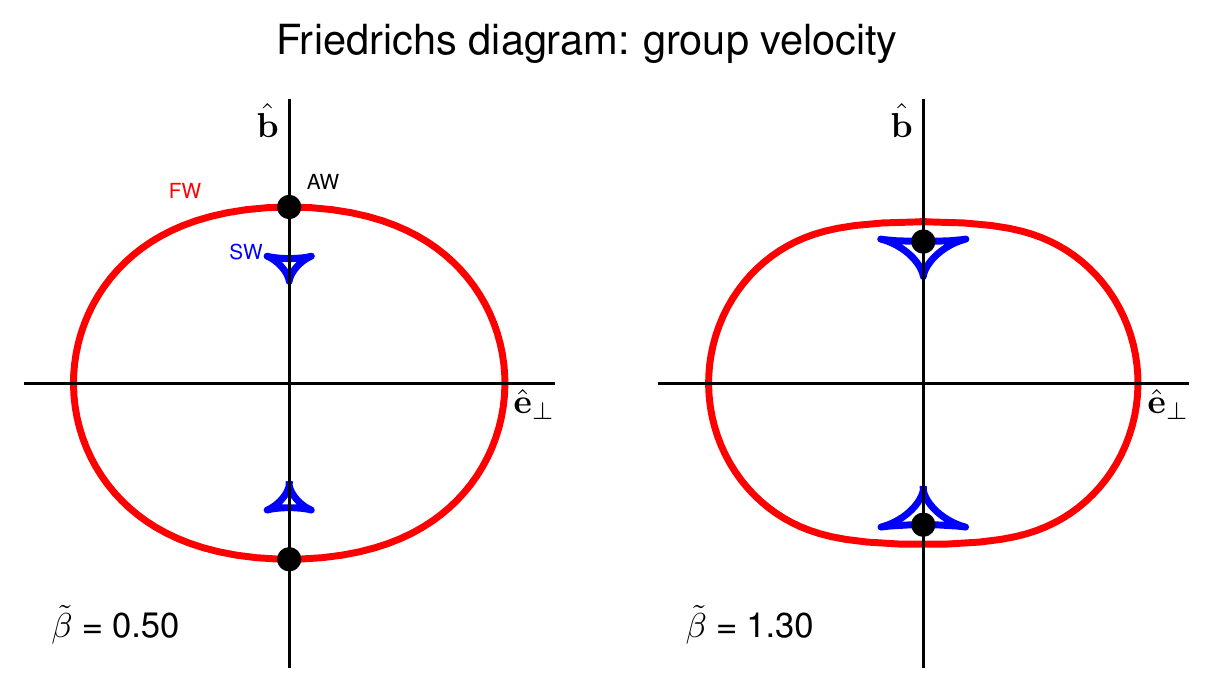}
    \caption{Friedrichs diagram for the group velocity of large-scale MHD waves. The layout is the same as in Fig. \ref{fig:PhaseDiag}. Note that the diagram for Alfvén waves consists in the two black dots due to the degeneracy of their group velocity.}
    \label{fig:GroupDiag}
\end{figure}

We immediately note an important result: the group velocity of pure AWs is degenerate as it does not depend on the wave vector ${\bf k}$. As a result any wave packet of pure AWs has to propagate along the magnetic field ${\bf \hat{b}}$. The same calculation can be followed for SWs and FWs, and one obtains

\begin{equation}
    \label{eq:GSaws}
    {\bf v}_g^{\pm} = v_p^{\pm} \left[ \hat{\bf k} \pm \frac{\sigma\cos\theta\sin\theta}{2\sqrt{1-\sigma\cos^2\theta}\left[1\pm\sqrt{1-\sigma\cos^2\theta}\right]}\hat{\bf n}  \right]\, ,
\end{equation}
where we have introduced $\sigma = 4\tilde{\beta}/\left(\tilde{\beta}+1\right)^2$, and $\hat{\bf n}$ is the direct unit vector perpendicular to $\hat{\bf k}$ in the $(\hat{\bf b}, \hat{\bf k})$ plane.

The group velocity indicates the propagation direction and speed of wave packets. It thus constitutes an important information to characterize how energy is channelled by wave ensembles. They are represented for AWs, SWs and FWs in Figure \ref{fig:GroupDiag} with the same layout as in Figure \ref{fig:PhaseDiag}. FWs are completely unfocused as their group velocity can take any direction. SWs have a more focused group velocity around  $\hat{\bf b}$. This focusing is nevertheless less and less tight as $\tilde{\beta}$ increases. Finally, AWs are completely focused (black dots in Figure \ref{fig:GroupDiag}), as we already noted earlier. This will have major consequences on the existence of Alfvén wings, which we now turn to.

\subsection{The concept of Alfv\'en wings}
\label{sec:AlfWings}

We will now apply the basic MHD concepts of Section \ref{sec:MHDwaves} to the case of star-planet magnetic interactions. We want to characterize the stationary waves that can exist due to the existence of an obstacle in a magnetized flow. In this simplified view, the obstacle can be a conducting planet, or the ensemble of a planet with its magnetosphere. The detailed shape of the obstacle itself does not matter for the moment. 

\begin{figure}
    \centering
    \includegraphics[width=\linewidth]{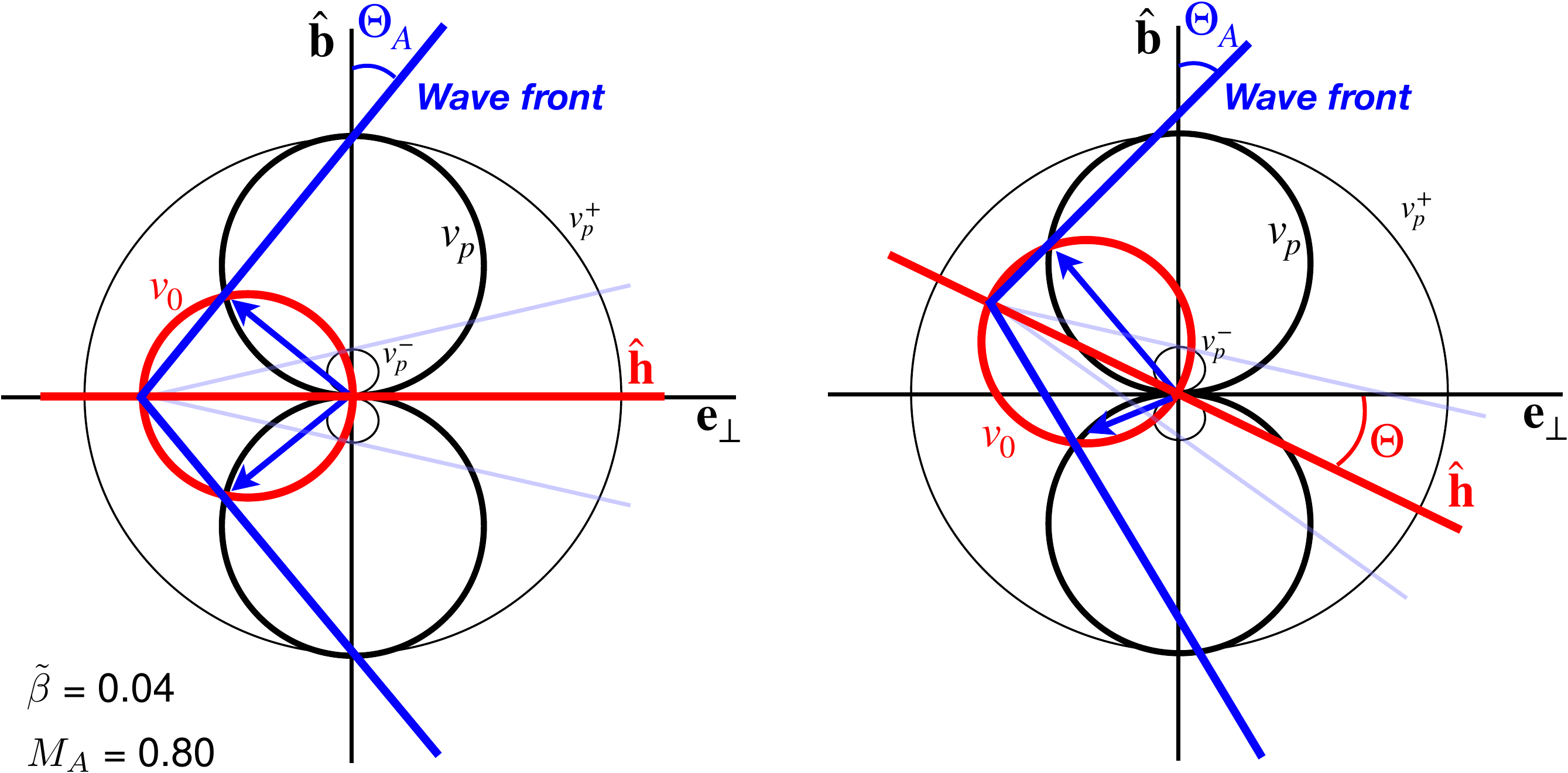}
    \caption{Friedrichs diagrams of the phase velocity involving a Doppler-shift associated with a large-scale flow impacting the obstacle. The Doppler-shift is labelled by the red circle. The direction of the incoming flow $\hat{\bf{h}}$ is given by the red line. The two panels show the same interaction case with a different impact angle $\Theta$ of the flow.}
    \label{fig:AWFdiag}
\end{figure}

We will consider the frame at rest with the obstacle. Let us define ${\bf v}_0$ the flow velocity in this frame, such that the obstacle is subject to a stream ${\bf v}_{\rm s} = v_0 \hat{\bf h}$ that deviates from being perpendicular to $\hat{\bf b}$ by an angle $\Theta$. In this frame, a stationary wave solution requires that
\begin{equation}
    \label{eq:Doppler}
    \omega = {\bf k} \cdot {\bf v}_0\, ,
\end{equation}
that is naturally the Doppler shift in this frame. If on applies the same reasoning for Doppler shift that the one for Alfv\'en waves developed in previous section, we remark that such waves are described in the Friedrichs diagram by a circle of radius $v_0/2$ centered on $(-\cos\Theta \, v_0/2,-\sin\Theta\, v_0/2)$ in the $(\hat{\bf b},{\bf e}_\perp)$ plane (again, in 3D this is actually a sphere by rotation along $\hat{\bf h}$). We can intersect this circle with the three Alfvén waves characteristics in the phase velocity Friedrichs diagram as shown in Figure \ref{fig:AWFdiag}. The possible stationary waves then depend on the most important control parameter for star-planet magnetic interactions: the (total) Alfvénic Mach number, defined here as

\begin{equation}
    \label{eq:AlfMachNumber}
    M_A = \frac{v_0}{v_A}\, .
\end{equation}

On the left panel of Figure \ref{fig:AWFdiag}, we consider the canonical case $\Theta = 0$ for a sub-alfvénic interaction ($M_A=0.8$). On the right panel, we illustrate the effect of the flow inclination angle $\Theta$. The Doppler-shift is labelled by the red circle. We rightfully note that it actually intersect only the characteristics of the SWs and AWs. If the interaction was super-alfvénic ($M_A>1$), the red circle would actually intersect the three characteristics. The Doppler, SW and AW circles all cross in the center. We find again the 'entropy wave', for which the wave vector ${\bf k}$ is perpendicular to ${\bf v}_0$ and the group velocity is equal to ${\bf v}_0$. These waves form a queue behind the obstacle, in the direction of the impacting flow.

We have seen in Section \ref{sec:MHDwaves} that the group velocity of AWs is degenerate. As a result, thanks to the Friedrichs diagram analysis we know that pure AW wave packets will propagate along the so-called Alfvén characteristics (blue lines in Figure \ref{fig:AWFdiag}) defined as 

\begin{equation}
    \label{eq:AWchar}
    {\bf c}_A^\pm = {\bf v}_0 \pm {\bf v}_A\, .
\end{equation}

Finally, the case of SWs is interesting. Alike the pure AWs, they show an intersection with the Doppler shifted red circle in Figure \ref{fig:AWFdiag}. Nevertheless, their group velocity is non-degenerate. As a result, any excited SW wave packets will occupy a larger and larger volume as we move away from the obstacle. The energy and momentum transport by these waves will therefore be unfocused and are generally ignored in the context of star-planet or planet-satellite interactions.  

\begin{figure}
    \centering
    \includegraphics[width=\linewidth]{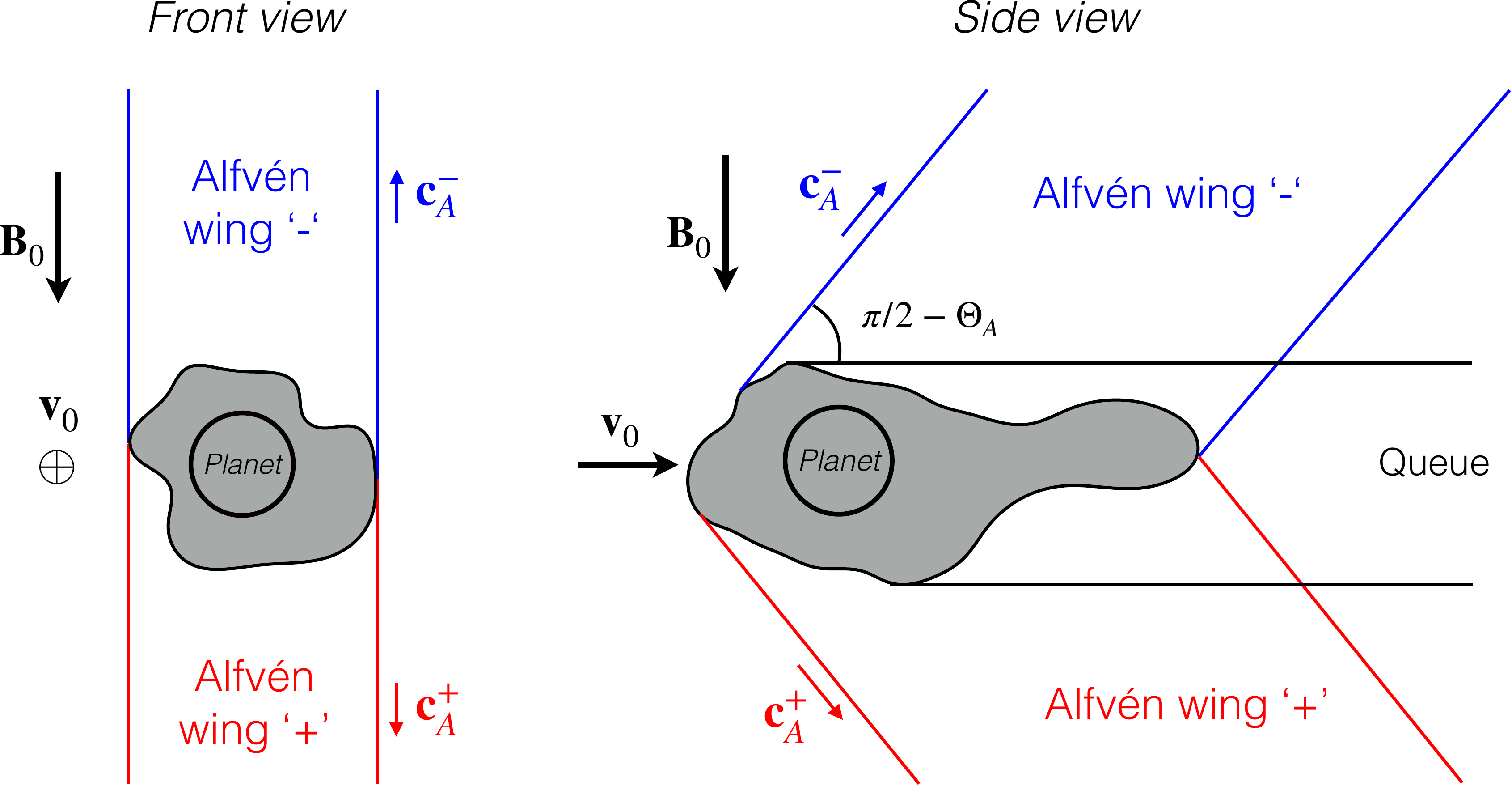}
    \caption{Schematic of the concept of Alfvén wings. The obstacle, illustrated by the blackened areas, can consist of a planet and its hypothetical magnetosphere. A front (left panel) and side (right panel) are shown. The two wings are labelled in red and blue.}
    \label{fig:AWschematic}
\end{figure}

We can sketch the characteristics of sub-alfvénic star-planet interaction deduced from this simple Friedrichs diagram analysis, as shown in Figure \ref{fig:AWschematic}. The Alfvén wings always live in the $(\hat{\bf b},\hat{\bf h})$ plane. The inclination angle between the Alfvén wings and the ambient magnetic field, $\Theta_A$, can be geometrically derived from Figure \ref{fig:AWFdiag} to obtain
\begin{equation}
    \label{eq:AWangle}
    \sin \Theta_A = \frac{M_A\cos\Theta}{\sqrt{1+M_A^2-2M_A\sin\Theta}}\, .
\end{equation}

To summarize, a first analytical description of Alfvén wings yields the following two robust conclusions:
\begin{itemize}
    \item Only pure Alfvén waves and slow MHD waves are excited by the obstacle in the subsonic flow. Between the two, only pure Alfvén waves packets propagate in a focused manner, which leads us to consider only this wave population in the energetic discussion hereafter (\S\ref{sec:Energetics}) 
    \item The overall shape of the Alfvén wings \textbf{does not depend on the specifics of the obstacle}. It is fully determined by the flow impacting the obstacle and the ambient magnetic field in the flow. 
\end{itemize}

\begin{figure}
    \centering
    \includegraphics[width=\linewidth]{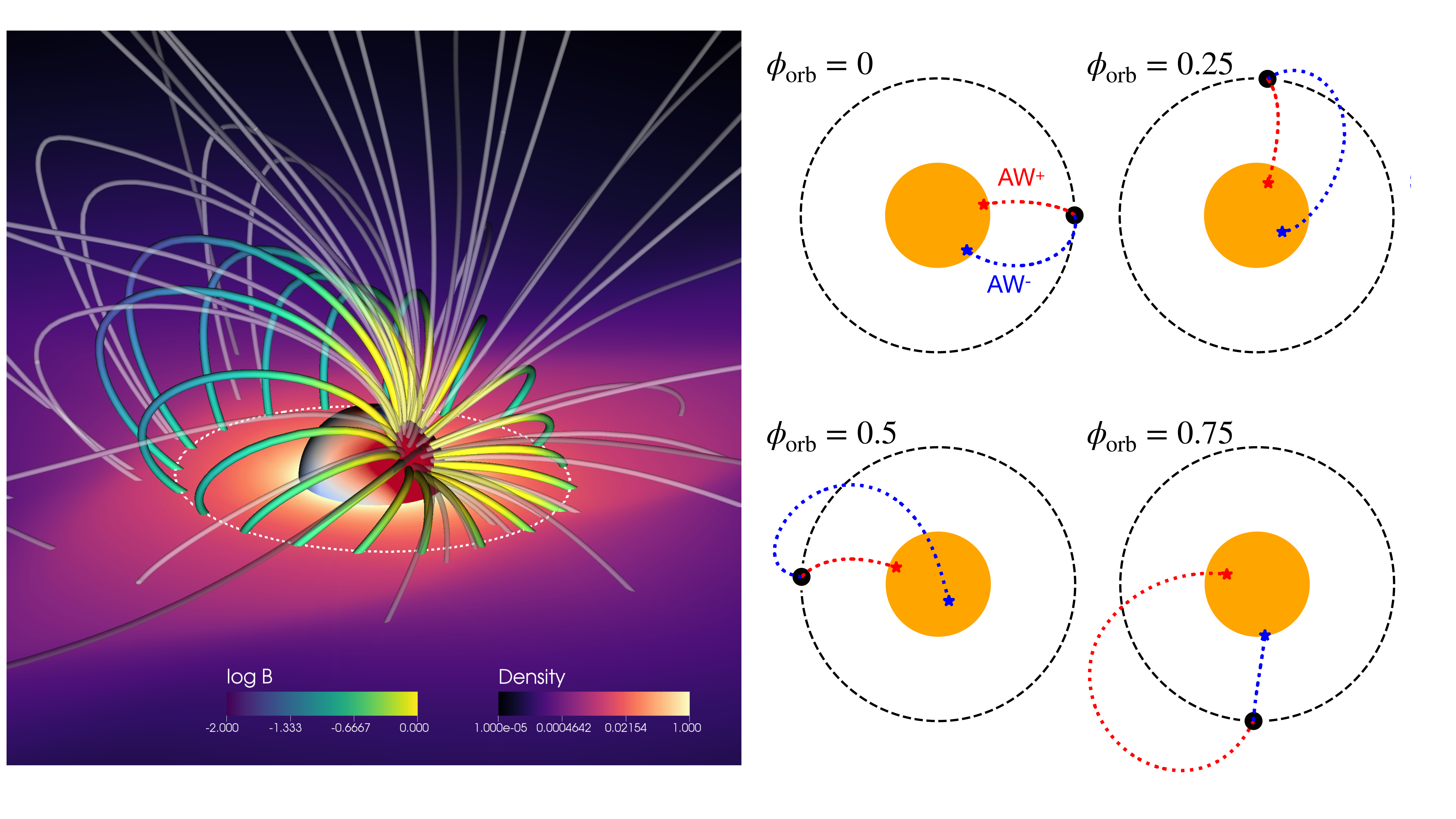}
    \caption{Alfvén wings in the Kepler-78 system \citep{Strugarek2019a}. The left panel is an illustrative view of the corona of Kepler-78 deduced from a 3D, MHD modelling using the observed magnetic map of Kepler-78 \citep{Moutou2016}. The right panels show the projection of the two Alfvén wings on the ecliptic plane for four different orbital phases $\phi_{\rm orb}$. Figures are adapted from \citet{Strugarek2019a} with permission of the AAS.}
    \label{fig:AWk78}
\end{figure}

The latter has a particularly interesting consequence: one can predict the shape of Alfvén wings solely based on the structure of a stellar corona and the orbital motion of an exoplanet. We show the application of the Alfvén wings theory to the Kepler-78 exosystem in Figure \ref{fig:AWk78} \citep[for more details see][]{Strugarek2019a}. On the left, a polytropic wind model is used to derive the plasma properties of the corona of Kepler-78, based on observed Zeeman-Doppler-Imaging magnetic map observed by \citet{Moutou2016}. The orbit of Kepler-78b is shown by the dashed white circle. On the right, we apply the Alfvén wings theory to predict how wings change along the orbit of the planet. We show the projected trace of north (in red) and south (in blue) wings on the ecliptic plane. We observe that as the planet orbit the AW shape and sub-stellar points (red and blue star symbols) dramatically change due to changes in magnetic connectivity in the stellar corona.  

\subsection{Energetics of Alfv\'en wings}
\label{sec:Energetics}

Now that we have understood how to characterize the shape of Alfvén wings, let us turn to their amplitude. How much energy is involved in magnetized star-planet interactions? The energy flowing through the wings is characterized by the Poynting flux $\bf{S}$ defined as 

\begin{equation}
    {\bf S} = \frac{{\bf E}\times{\bf B}}{\mu_0}\, . 
\end{equation}

Assuming ideal magneto-hydrodynamics, the electric field is simply ${\bf E} = - {\bf v}\times{\bf B}$. The projection of ${\bf S}$ along the Alfv\'en characteristics in the plasma reference frame is therfore

\begin{equation}
    S_{AW} = {\bf S} \cdot \frac{{\bf c}_A^\pm}{\left|{\bf c}_A^\pm\right|} = \frac{v_0 B_0^2}{\mu_0} \sin\Theta_A \cos\Theta \, .
    \label{eq:PoyFluxBasic}
\end{equation}

Equation \ref{eq:PoyFluxBasic} gives a robust, maximum estimate of the power involved in the magnetic interaction. It corresponds to the case where the perturbations propagating in the wings are sufficiently large so that the incoming flow is completely halted inside the wings. 

\begin{figure}
    \centering
    \includegraphics[width=\linewidth]{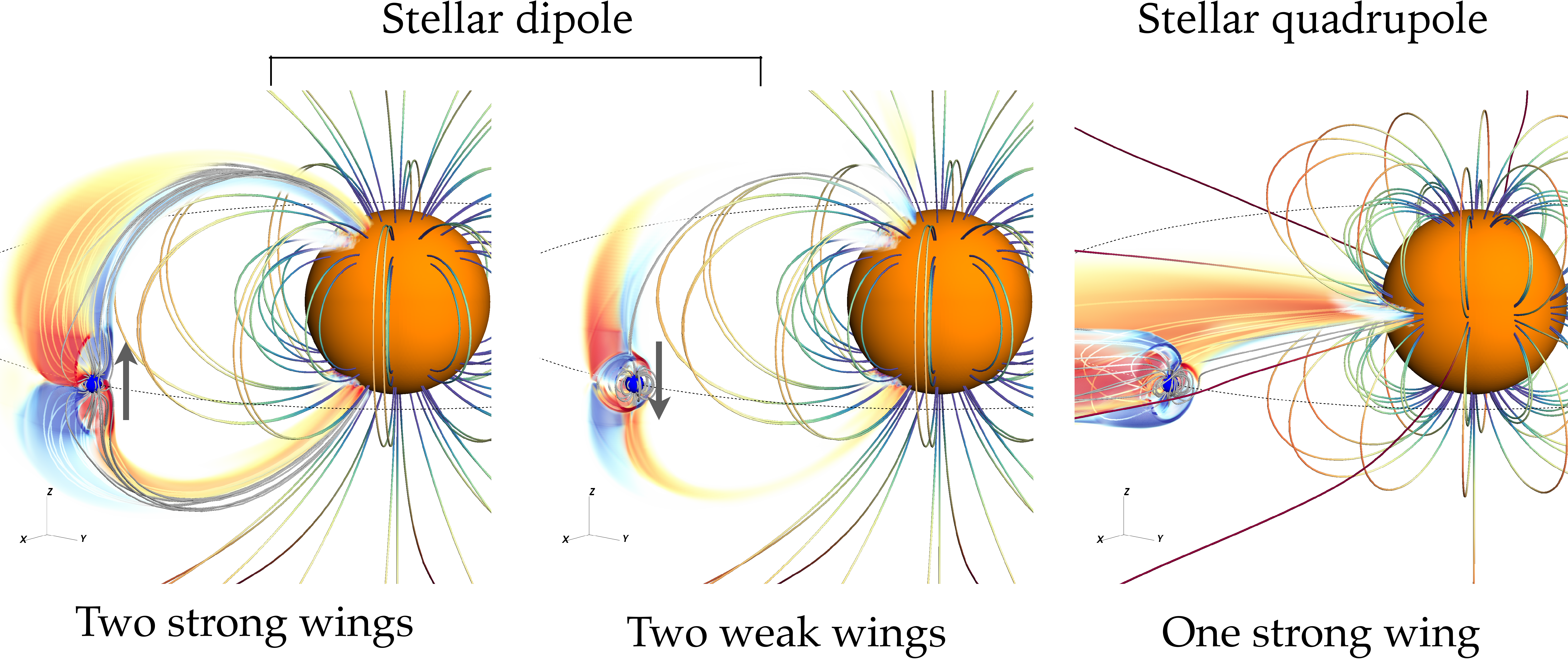}
    \caption{Illustration of the effect of the magnetic topology on Alfvén wings, from the 3D numerical simulations of \citet{Strugarek2015}. The two left cases are the same except for the orientation of the planetary field (grey arrow), showing a modulation of the wings in \textbf{amplitude} (the red/blue volume renderings show the current system of the Alfvén wings). The right panel is the same except for the stellar magnetic field that is quadrupolar, showing a change in the \textbf{shape} of Alfvén wings. Figures are adapted from \citet{Strugarek2015} with permission of the AAS.}
    \label{fig:Topology}
\end{figure}

In general, the exact Poynting flux depends on two important aspects of star-planet magnetic interaction: (i) the details of the interaction inside the grey area in Fig. \ref{fig:AWschematic}: what are the plasma, magnetic, and conductive properties of the obstacle? and (ii) the size of the effective obstacle. The latter determines the thickness of the Alfvén wings, over which one has to sum to estimate the total power involved in the interaction. We illustrate these aspects in Fig. \ref{fig:Topology} where we show 3D simulations for three different magnetic topologies \citep{Strugarek2015}. The Alfvén wings are highlighted by the volume rendering of the currents in blue/red (negative/positive). All three cases are equivalent: in the two first cases only the orientation of the planetary field is reversed, while in the third case the stellar magnetic field is changed from a dipole to a quadrupole of similar amplitude. We strikingly see in these renderings that the relative topology of the magnetic field has an important influence on the strength of the wings, while their overall shape is control by the stellar atmosphere. 

\citet{Saur2013} have developed an analytical model to calculate the total Poynting flux over Alfvén wings, assuming the obstacle is magnetized. They found, for simple geometries, 

\begin{equation}
  \label{eq:PoyFlux}
    \mathcal{P} = \int_{AW} {\bf S} \cdot {\rm d}{\bf A} \simeq 2\pi R^2 \bar{\alpha}^2 \left(1+M_A^2 - 2M_A\sin\Theta\right)^{1/2} S_{AW}\, ,  
\end{equation}

where $R$ is the effective radius of the obstacle, and $\bar{\alpha}$ quantifies the conductive properties of the obstacle\footnote{$\bar{\alpha}=\Sigma_P/(\Sigma_P+\Sigma_A)$, where $\Sigma_P$ is the Pedersen conductance in the planet's ionosphere and $\Sigma_A=(1/\mu_0v_A)(1+M_A^2-2M_A\sin\Theta)^{-1/2} $ is the Alfvén conductance.}. These results were later confirmed and extended to more complex geometries through 3D numerical simulation in \citet{Strugarek2016c,Strugarek2017e}. The maximal power is obtained for high conductance in planet's atmopsheres for which $\bar{\alpha}=1$. In the solar system, $\bar{\alpha}$ varies from a few $10^{-2}$ to about 1 for the different moons of Jupiter and Saturn \citep{Kivelson2004,Saur2013}. The conductive properties of exoplanets are nevertheless largely unknown as of today. They depend for instance on the complex photo-chemical reactions in their upper atmosphere, and hence on both the irradiation recevied by the planet, and the planet atmospheric composition. Both are expected to change significantly along stellar evolution. Ab-initio calculations are today being performed to characterize (among other aspects) the conductive properties of exoplanets' atmosphere (\textit{e.g.} see the Kompot code, \citealt{Johnstone2019}). 

\begin{figure}
    \centering
    \includegraphics[width=\linewidth]{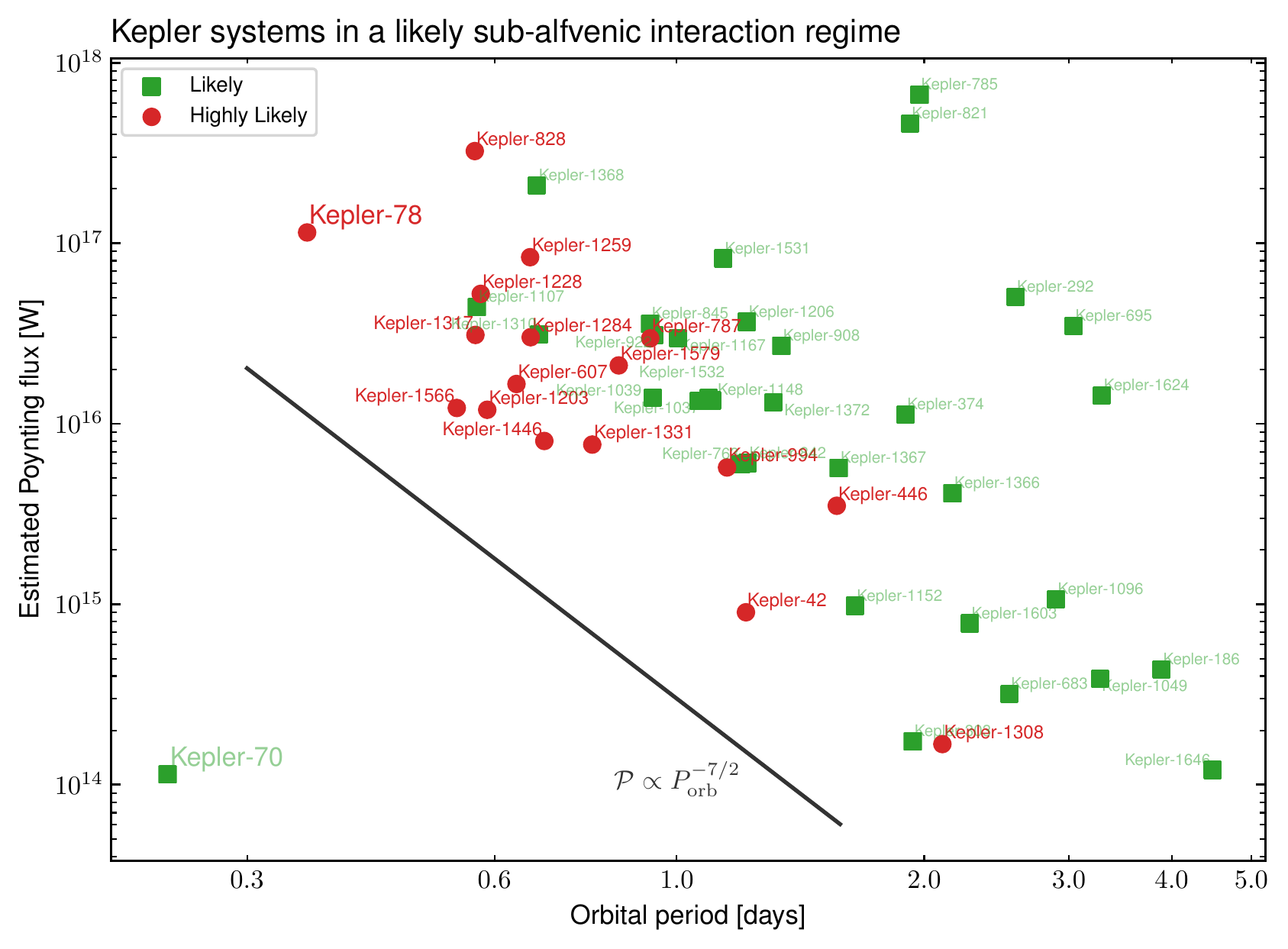}
    \caption{Estimated Poynting flux power (Eq. \ref{eq:PoyFlux}) for the Kepler sub-sample of \citet{McQuillan2013} as a function of the orbital period of the planet. The red dots label the planets that are always predicted to exhibit star-planet magnetic interactions within all the error-bars of the model. The green squares label the planets that are predicted to exhibit such interactions only for the most optimistic error-bars.}
    \label{fig:PoyKepler}
\end{figure}

Assuming a maximized interaction, the total Poynting flux $\mathcal{P}$ requires the knowledge of the plasma environment around a planet along its orbit. \citet{Saur2013} considered a simple Parker-like wind and applied their formalism to the ensemble of known exoplanets at that time. The largest uncertainty lies here in the lack of knowledge of the key stellar properties (rotation period, magnetic field) in the sample. For this review I revisited the updated ensemble of \href{http://kepler.nasa.gov}{Kepler}-observed exosystems while using the wind-modelling strategy developed by \citet{Ahuir2020}. This strategy can be applied for the stars with a detected rotational period \citep{McQuillan2014}: a stellar Rossby number can then be estimated based on the scaling law derived from 3D numerical simulation by \citet{Brun2017a}. The stellar magnetic field and the subsequent wind is then computed following the approach proposed by \citet{Ahuir2020} that satisfies all the observational contraints on stellar wind available to date. The resulting estimated Poynting flux (Eq. \ref{eq:PoyFlux}) assuming a planetary surface field of 4G for these systems is shown in Fig. \ref{fig:PoyKepler}. In spite of the large scatter due to the large variety of star-planet systems considered here, we see a clear overall trend
$  \mathcal{P} \propto P_{\rm orb}^{-7/2}$
for the compact exosystems in the \href{http://kepler.nasa.gov}{Kepler} field. We highlight two populations in Fig. \ref{fig:PoyKepler}: exosystems in a very likely sub-aflvénic interaction regime (red circles), and exoplanets in a likely sub-alfvénic interaction regime (green squares). The former population is found to be in a sub-alfvénic interaction within all the error-bars of the models and observations used here (see \citealt{Ahuir2020} for more details on these error-bars). The latter population (green squares) is predicted to be sub-alfvénic only for part of our estimated magnetic field and coronal properties of the central star. It is nonetheless instructive to observe that sub-alfvénic interactions are highly unlikely to occur for planets with an orbital period longer than five days. 

The Kepler-70 system stands out in this diagram. It is composed of a hot sub-dwarf around which a super-Earth (Kepler-70b) orbit with an extremely short period. Our estimates predict a moderate stellar magnetic field, which translates into a quite weak star-planet magnetic interaction. As a result, small orbital periods do not necessarily imply strong star-planet magnetic interactions. This is also illustrated by the large vertical spread in Fig. \ref{fig:PoyKepler}: planets with the same orbital period can lead to absolute Poynting fluxes varying by more than 3 orders of magnitude (\textit{e.g.} for $P_{\rm orb} \sim 2$ days) depending on the spectral, rotational, and magnetic properties of their host star.

The case of Kepler-78 is also highlighted in the top left of Figure \ref{fig:PoyKepler}. The Poynting flux estimate was carried out using our empirical scaling law with no \textit{a priori} knowledge of the stellar magnetic field. It nonetheless provides an estimate that is very close to the one obtained when carrying out fully 3D MHD simulations of Kepler-78, based on an observed magnetic maps with Zeeman-Doppler Imaging (see Fig. \ref{fig:AWk78} and \citealt{Strugarek2019a}).

We have illustrated the theoretical derivation of the energetics of star-planet magnetic interaction with the \href{http://kepler.nasa.gov}{Kepler} sub-sample of exoplanets. This sub-sample is particularly interesting because asterosismology allows us to have access with good precision to the fundamental stellar parameters required to properly assess magnetic interactions. Many other exosystems as for instance HD 1897333 or 55 Cnc \citep{Cauley2019} also present promising configurations for sub-alfvénic interactions. Dedicated, 3D models of such systems are today required to properly assess both the energetics and the ephemerids of star-planet interactions \citep{Strugarek2019a}. One can hope that the coming years will see the definitive confirmation of star-planet magnetic interactions detection thanks to a combination of advanced 3D models supporting this interpretation of peculiar observational features correlated with exoplanets' orbital periods.

\subsection{Angular momentum transfer and the secular evolution of stellar systems}
\label{sec:Amom}

The mere existence of Alfvén wings implies an energy transfer that also contains in part  an \textit{angular momentum transfer}. Indeed, an orbiting planet interacts with its environment and therefore is subject to an equivalent drag force. For the solar-system planets, the stellar wind is not quite dense enough for this force to matter at all in their orbital evolution. But it may not be systematically the case for close-in exoplanets orbiting in a dense and strongly magnetized stellar coronae. 

\begin{figure}
    \centering
    \includegraphics[width=\linewidth]{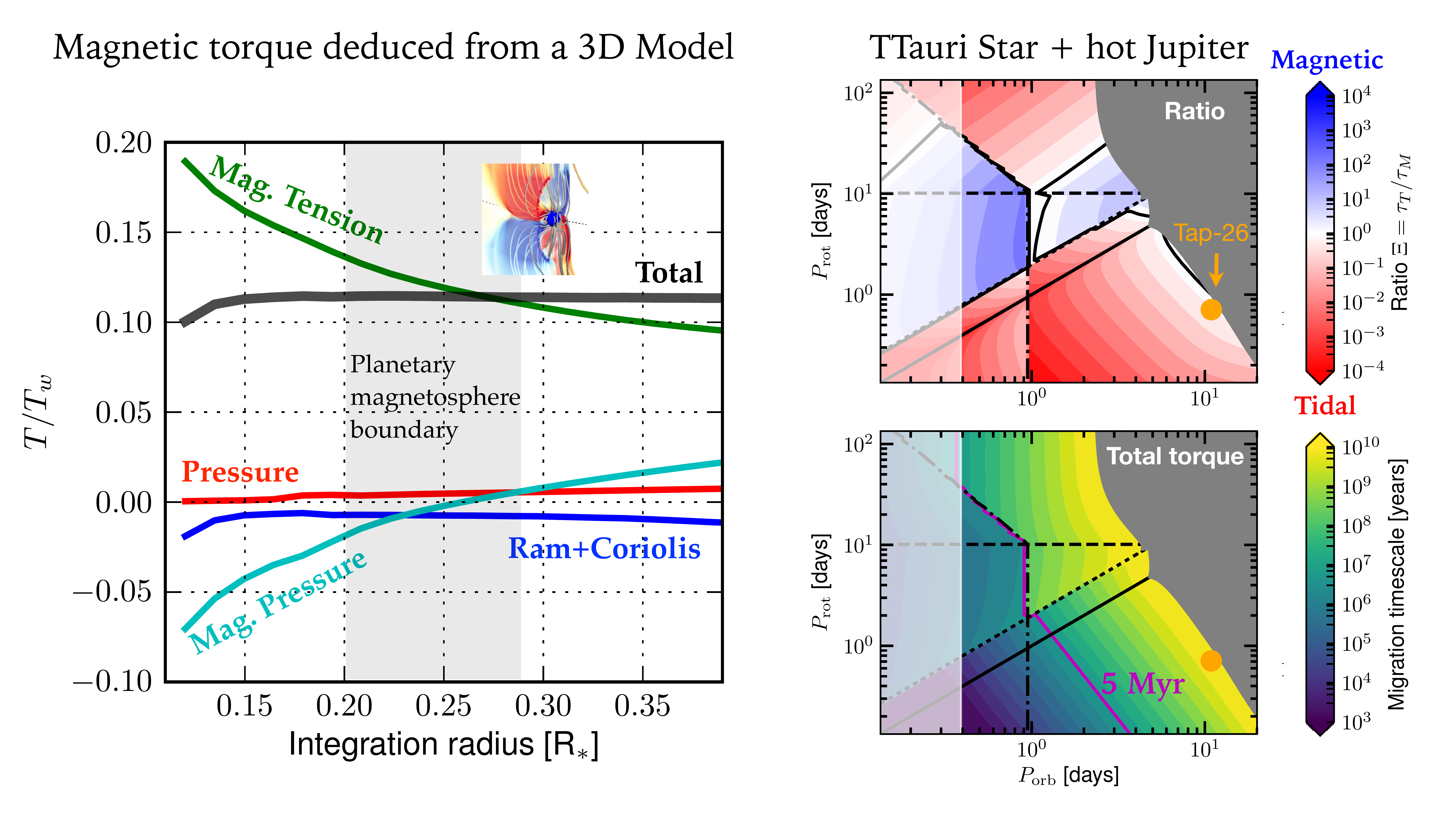}
    \caption{Illustration of magnetic torques due to star-planet magnetic interactions. The left panel shows an angular momentum budget around the orbiting exoplanet from 3D numerical simulations \citep{Strugarek2015}. The main magnetic torque is dominated by the magnetic tension contribution in green. The right panel shows the comparison between tidal and magnetic migration torques in a ($P_{\rm rot},P_{\rm orb}$) diagram for TT-Tauri -- hot Jupiter exosystem (see \citealt{Strugarek2017c} for more details). The bottom right panel shows the total (tidal+magnetic) migration timescale in a logarithmic color-scale for the same system. Figures are adapted from \citet{Strugarek2015} and \citealt{Strugarek2017} with permission of the AAS.}
    \label{fig:Torques}
\end{figure}

The magnetic torque felt by close-in exoplanets was systematically parameterized through numerical simulations in \citet{Strugarek2016c,Strugarek2017e}. We show the example of one such simulation in the left panel of Fig. \ref{fig:Torques}, for one particular magnetic topology (see \citealt{Ip2004,Strugarek2015} for more details on the effect of topolgy). A clear magnetic torque originating from the magnetic tension around the planet (green line) produces most of the total net torque (grey line) felt by the orbiting planet. The main unexpected finding of these studies was that this net magnetic torque could make close-in planets migrate on a time-scale of hundreds of millions years. Henceforth, magnetic torques could in principle play a role in shaping the population of close-in exoplanets. This situation typically arises mainly for strongly magnetized dwarf stars, or early in the evolution of a solar-like star. For main-sequence slow-rotating stars, magnetic torques play a more negligible role: for instance, the fastest migration timescale found in the \href{http://kepler.nasa.gov}{Kepler} sub-sample studied in Fig. \ref{fig:PoyKepler} is predicted for Kepler-828 with a timescale of approximately 2 Gyrs.

Nevertheless, it is important to consider magnetic torques for ensemble evolution of compact exosystems, particularly during their early evolution. A systematic comparison between tidal and magnetic torques can be carried out thanks to the developments in both theoretical fields over the last decade. We have carried out such a comparison in \citet{Strugarek2017c} and showed that in general both physical phenomenons add up to make orbital migration more efficient. We found that magnetic torques were likely dominating the migration process for fully-convective stars and young stars, while tidal torques were likely dominating otherwise. We summarize this instantaneous comparison in the right panel of Fig. \ref{fig:Torques}, for a typical TTauri star - hot Jupiter system (alike Tap-26, see \citealt{Yu2017}). We see that the torque can be dominated by either tidal torques (red regions in the top panel) or magnetic torques (blue regions) in a $P_{\rm rot}$-$P_{\rm orb}$ diagram. The bottom panel shows the migration timescale resulting from the sum of tidal and magnetic torques (see \citealt{Strugarek2017c} for more details).

It is today mandatory to go beyond the instantaneous approach to understand how compact star-planet systems jointly evolve. Any orbital migration corresponds to an angular momentum exchange with the star, that ultimately affects the stellar rotation period, and thus its dynamo action and the sub-sequent evolution of the system. This is a challenging endeavour as it involves the complex stellar secular evolution along with non-linear physical phenomena such as stellar winds, and tidal and magnetic star-planet interactions. A seminal exploration was carried out simultaneously by \citet{Bolmont2012,Penev2012,Zhang2014} exploring the angular momentum budget of a two-body star-planet system along stellar evolution, including tidal interactions. Recently, \citet{Gallet2018} followed a similar approach based on state-of-the-art stellar evolution tracks produced by the STAREVOL code \citep{Amard2016a}, and advanced stellar wind \citep{Matt2012a} and tidal interactions \citep{Ogilvie2013,Bolmont2016} modelling. In this study they focused on the hot exoplanets population predicted by such models for young open clusters such as Pleiades. A similar approach was also used with the ESPEM code \citep{Benbakoura2019} to predict which star-planet systems were the most likely to undergo a coupled orbital-rotational evolution phase. Star-planet magnetic torques yet have to be included in these theoretical explorations. This tool can ultimately be used to produce synthetic exoplanet populations, as we recently demonstrated in \citet{Ahuir2021}.

Such studies are today highly valuable in the context of \href{http://kepler.nasa.gov}{Kepler} and \href{https://tess.mit.edu}{TESS} (and in the future \href{https://www.cosmos.esa.int/web/plato}{PLATO}) as the number of the observed exosystems keeps growing fast. They will allow to disentangle the relative importance of the different physical effects in shaping the population of exoplanets thanks to dedicated, careful cross-comparisons between observed and synthetic samples. 

\section{Conclusions}
\label{sec:conclusions}

We have presented here the basics behind our physical understanding of star-planet magnetic interactions. Several \textit{grey} areas nevertheless remain to be explored and understood. Among them, I would like to highlight in particular that we have almost not discussed the details of the obstacle itself, \textit{i.e.} the planet and its hypothetical magnetosphere/ionosphere/etc... While the toplogy of star-planet magnetic interactions is largely controlled by the star itself, their strength is sensitive to the detailed properties of the obstacle. What are the conductivites in the different layers of the atmosphere and interior of exoplanets? This question cannot be trivially anserwed today, as exoplanets can have wildly different chemical reactions compared to solar system planets, thay can receive a much different ionizing flux from their star, and they can be subject to much stronger tidal interactions with their host star and/or other planets in the system.

In spite of these hurdles that have to be overcome on the theoretical side, the characterization of star-planet magnetic interactions should allow us to probe the magnetism of exoplanets. This grand goal is worth such efforts, as today we can only characterize (with not so much precision) the magnetism of the few planetary bodies in the solar system. One may await a revolution in our understanding of planetary magnetism if the hunt for star-planet magnetic interactions succeed in the future. I hope this review will help motivate more and more young researchers to dive into the fantastic open world of star-planet interactions.

\acknowledgements

I warmly thank J. Saur and K. Khurana for useful discussions on planet-satellite and star-planet interactions, and the exoplanet Team behind the \href{http://exoplanet.eu}{exoplanet.eu} website. I want to thank also J. Ahuir for a careful read of this manuscript. I am grateful to INSU/PNP for its financial support for this research. The numerical simulations shown in this manuscript were made possible thanks to GENCI time allocations over the years.


\end{document}